\documentclass[pre,floatfix,twocolumn]{revtex4}
\usepackage{latexsym}
\usepackage{amsmath, amsthm, amssymb}
\usepackage{mathrsfs}
\usepackage{graphicx}
\usepackage{epstopdf}
\usepackage{dcolumn}
\usepackage{expl3}
\usepackage{float}
\usepackage{hyperref}
\usepackage{placeins}
\usepackage{setspace}
\setcitestyle{square,numbers}
\allowdisplaybreaks

\usepackage{boldline}
\usepackage{enumerate}
\usepackage{wrapfig}
\usepackage{multirow}
\usepackage{array}
\usepackage{makecell}
\usepackage{longtable}
\usepackage{verbatim}

\makeatletter
\renewcommand*{\thesection}{\arabic{section}}
\renewcommand*{\thesubsection}{\thesection.\arabic{subsection}}
\renewcommand*{\p@subsection}{}

\renewcommand*{\p@subsubsection}{}
\makeatother

\begin{document}

\title{The interplay of inhibitory and electrical synapses results in complex persistent activity}

% Use letters for affiliations, numbers to show equal authorship (if applicable) and to indicate the corresponding author
\author{R.~Janaki}
\email{janaki.phys@gmail.com}
\affiliation{Department of Theoretical Physics, University of Madras, Maraimalai Campus,
Chennai, Tamil Nadu-600025, India}
\affiliation{The Institute of Mathematical Sciences, CIT Campus, 
Taramani, Chennai 600113, India}

\author{A.~S.~Vytheeswaran}
%\email{sitabhra@imsc.res.in}
\affiliation{Department of Theoretical Physics, University of Madras, Maraimalai Campus,
Chennai, Tamil Nadu-600025, India}

\date{\today}

\begin{abstract}
Inhibitory neurons play a crucial role in maintaining persistent neuronal activity. Although connected extensively through electrical synapses (gap-junctions), these neurons also exhibit interactions through chemical synapses in certain regions of the brain. When the coupling is sufficiently strong, the effects of these two synaptic modalities combine in a nonlinear way. Hence, in this work, we focus on the strong inhibition regime and identify the parametric conditions that result in the emergence of self-sustained oscillations in systems of coupled excitable neurons, in the presence of a brief sub-threshold stimulus. Our investigation on the dynamics in a minimal network of two neurons reveals a rich set of dynamical behaviors viz., periodic and various complex oscillations including period-$n$ ($n=2,4,8\ldots$) dynamics and chaos. We further extend our study by considering a system of inhibitory neurons arranged in a one-dimensional ring topology and determine the optimal conditions for sustained activity. Our work highlights the nonlinear dynamical behavior arising due to the combined effects of gap-junctions and strong synaptic inhibition, which can have potential implications in maintaining robust memory patterns.
\end{abstract}

\maketitle
%\linenumbers
%\section*{Keywords}
%Excitable, Coupled neurons, Inhibitory synapse, Gap-junctions, nonlinear %dynamics, Complex oscillations, Chaos, Persistent activity

\section{Introduction}
\paragraph*{}An intriguing aspect of populations of coupled neurons is their ability to generate persistent activity~\cite{Major2004}. Such sustained activity is found in diverse brain areas~\cite{Schultz2003, Komura2001, Prut1999, Moschovakis1997, Kojima1996} and species~\cite{Marques2020, MajorA2004} and the qualitative similarity in their activity patterns %in all different organisms 
indicate their importance and the possibility of an unified mechanism underlying such brain dynamics. Several studies suggest a potential link between persistent activity and working memory of the brain~\cite{Wang1999, Compte2000, Zylberberg2017, Kaminski2017}, which is crucial for many cognitive processes including decision-making~\cite{Curtis2010, Haller2018}. This remarkable ability of the brain to achieve stable persistent state which in turn enables robust information storage~\cite{Curtis2003} is attributed to the efficient communication between its constituent elements which are the neurons. It is well known that the inter-neuronal communication in the mammalian brain is largely achieved by chemical synapses~\cite{Galarreta2001}. Such synapses can either be excitatory or inhibitory (depending on the neurotransmitter they release) and the precise balance between them is crucial for proper brain functioning~\cite{Symonds1959, Rubenstein2003, Uhlhaas2010, Yizhar2011}. Nevertheless, the role of electrical synapses (or gap-junctions) in maintaining normal physiological function and homeostasis cannot be ignored~\cite{Dong2018}. Hence, it is important to understand the interplay between these two synaptic modalities viz., chemical and electrical in generating persistent neuronal activity.

Networks of inhibitory neurons have been studied extensively in the context of synchronization and are known to play a key role in generating robust network oscillations~\cite{Buzsaki2012, Allen2015}. Although they constitute only 10\%-20\% of the neuronal population~\cite{Buzsaki2007, Swanson2019}, they play a significant role in sculpting the network dynamics. These inhibitory neurons are known to be connected predominantly through electrical gap-junctions ~\cite{Galarreta2001}, in addition to their synaptic connections. Previous studies on coupled inhibitory neurons focused on studying synchronization among the constituent neurons with gap-junctions alone~\cite{Chow2000, Lewis2003A} and with synapse and gap-junctions~\cite{Lewis2003, Kopell2004, Pfeuty2005}. Thus it is apparent from these studies that gap-junctions play a crucial role in enabling neuronal synchronization. This tendency of gap-junctions to promote synchronization could be one of the prominent reasons for their lack of occurrence between excitatory neurons, as synchronization prevents persistent activity~\cite{Ermentrout2006}. As the present work focuses on mechanisms underlying persistent activity as opposed to the previously studied neuronal synchronization, considering networks that only comprise inhibitory neurons can enhance our understanding towards that direction.

Chaotic dynamics is known to exist in many biological systems ranging from coupled genetic circuits~\cite{Kappler2003, Novak2008,  Zhang2012, Suzuki2016, Liu2019}, cardiac cells~\cite{Glass1983, Chialvo1990, Lewis1990, Garfinkel1997} and ecological networks~\cite{Schaffer1985, Hastings1991, Hassell1991, Upadhyay1998, Upadhyay2009, Pearce2020} and there has been plenty of research of such systems reporting chaos with and without delays. Investigations of the dynamics of individual neurons in the network coupled through synapses, have revealed that the activity profile does not necessarily exhibit rhythmic behavior. In particular, networks of synaptically coupled neurons are shown to exhibit irregular dynamical activity, often attributed to chaos~\cite{Jahnke2008, Shim2018}. There has been plenty of evidence for the  existence of non-periodic dynamics in a single neuron (e.g. non-periodic oscillations in internodal cell of Nitella flexillis~\cite{Hayashi1982},  chaotic oscillations in Molluscan neurone~\cite{Holden1982}). Such chaotic dynamics occurs not just at the level of single neurons but at several hierarchical levels in the brain~\cite{Amari2003, Freeman2003, Korn2003}. Although the occurrence of chaos both during normal and pathological brain states suggest their significance in neuronal mechanisms, their precise functional role still remains unclear~\cite{Freeman1995}.  Nevertheless, it is compelling to see if chaos appears as an emergent persistent behavior in a network comprising of only inhibitory neurons in the presence of both chemical and electrical synaptic coupling.

In this paper we address the following questions: (1) how does the interplay between inhibitory synaptic and gap-junctional (electrical) coupling result in persistent activity? (2) what is the simplest network that can generate complex sustained dynamics? To this end, we first analyze in detail the dynamical behavior exhibited by two coupled excitable neurons (with each neuron generating an action potential when a sufficient stimulus is applied), coupled through both an uni-directional synapse and a bi-directional gap-junction. By applying a short pulse to one of the constituent neurons (specifically the pre-synaptic neuron), we investigate the conditions that result in persistent neuronal activity rather than previously studied synchronization, where the individual neurons were chosen to be oscillators and not excitable elements. We know that if the coupling strengths of both the synapse and the gap-junctions are sufficiently strong, they combine in a nonlinear manner and can give rise to new, complex behavior of the system~\cite{Pfeuty2005}. Motivated by a study by Kopell and Ermentrout [2004], for this work, we focus on strong synaptic coupling regime and study the effects of varied gap-junctional coupling strengths on generating persistent activity. As one of our key results, we demonstrate the emergence of complex dynamical behavior such as period-$n$ ($n=2,4,8\ldots$) oscillations and chaos using the simplest setting consisting of an uni-directional synapse from an inhibitory pre-synaptic neuron and a gap-junction, with the neurons starting from their resting state as well as from random initial conditions. We perform a detailed parametric study for systems in the presence of brief pulse and obtain parameter space diagrams that indicate the various attractors to which the system converges to when starting from resting state initial conditions, viz., no oscillations, periodic oscillations and complex oscillations. We further extend our study to a ring of inhibitory neurons having synapses between randomly chosen pairs of neurons and a bi-directional gap-junction with their nearest neighbors. Using this set up, we obtain the basin size of the various attractors in the system and also determine the optimal conditions for obtaining sustained activity under strong inhibition. Thus, in this work, we present a detailed picture of a minimalistic network of neurons coupled through synapses and gap-junctions. This not only enables deeper understanding of the mechanism uncovering persistent activity under strong inhibition but can also aid future research on addressing broader questions related to cortical computation and controlling memory patterns~\cite{Mongillo2018}.
%\textcolor{red}{Most of the previously reported studies of network of %neurons coupled through either inhibitory synapses and/or gap-junctions %focussed on the synchronization phenomena~\cite{Baptista2010}, which could %be related to pathological brain activity like epilepsy~\cite{Volman2011}.} 

%\subsection{`Review' option}
\section*{II. The Model} \label{sec:model}
\paragraph*{}We consider a system of $N$ identical Fitzhugh Nagumo (FHN) neurons~\cite{Fitzhugh1961, Nagumo1962}, coupled through chemical synapses and electrical gap-junctions. The dynamics of the coupled system is described by the equations:%Eq.~\ref{eq:FHN}
\begin{equation} 
\begin{split}
%\epsilon\dot{V_i} &= \frac{V_i\ (1 -V_i)\ (V_i - a) - W_i\ + I_{ext} - %I_i^{syn} + I_i^{gap}}{\epsilon},\\
\epsilon\dot{V_i} &= {V_i\ (1 -V_i)\ (V_i - a) - W_i\ + I_{ext} - I_i^{syn} + I_i^{gap}},\\
\dot{W_i} &= V_i - k\ W_i,
%\dot{s} &= tanh()
\end{split}
\label{eq:FHN}
\end{equation}
where $i \in \{1,2...N\}$ denotes the neuron index, $V_i$ is the associated membrane potential, $W_i$ is the associated recovery variable
and $I_{ext}$ is the external current. The parameters $a = 0.1 $ and $k = 0.5$ describe the model kinetics, while $\epsilon =0.01$ is the recovery rate. These values are chosen such that each uncoupled neuron is an excitable system.

	The $I_{i}^{syn}$ appearing in Eqn.~(\ref{eq:FHN}) represents the synaptic current which is modeled here as an ohmic current~\cite{Destexhe1994, Ermentrout2010}. Synapses in the brain are uni-directional with synaptic coupling from the pre-synaptic neuron $j$ to the post-synaptic neuron $i$. The equation for the synaptic current onto the post-synaptic neuron $i$ is given by:
\begin{equation}
\begin{split}
I_i^{syn} = g_{syn} \ \sum_{j=1}^{N}  A_{ij} \ (V_i- E_{syn})\ s_{ji},
\end{split} 
\end{equation}
where $g_{syn}$ is the synaptic conductance, $A_{ij}$ represents the synaptic weight matrix (or the adjacency matrix) and $E_{syn}$ represents the synaptic reversal potential with $E_{syn} = 5$ for excitatory and $E_{syn} = -5$ for inhibitory neurons respectively. $s_{ji}$ denotes the synaptic gating variable that evolves according to the equation,
\begin{equation}
\begin{split}
\dot{s}_{ji} &= \alpha\ N(V_j)\ (1-s_{ji})\ - \beta\ s_{ji},\\
\end{split} 
\label{eq:Gating}
\end{equation}
where,
\begin{equation*}
\begin{split}
N(V_j) &= 0.5\  (1+\tanh((V_j-v_{th})/v_{sl})).
\end{split} 
\end{equation*}
Here $\alpha=3$ and $\beta=3$ are the decay constants and $v_{th} = 0.3$, $v_{sl} = 0.001$ are the parameters that determine the shape of the synaptic term. The equation for the gating variable (\ref{eq:Gating}) thus depends only on the membrane potential of the pre-synaptic neuron $V_j$. 

  The $I_{i}^{gap}$ appearing in Eqn.~(\ref{eq:FHN}) represents the gap-junctional current. Such electrical coupling between the neurons is diffusive in nature and hence the gap-junctional current can be written as:
\begin{equation}
\begin{split}
I_i^{gap} = g_{gap} \ \sum_{nn} (V_j- V_i),
\end{split} 
\end{equation}
where $g_{gap}$ represents the gap-junctional conductance and the summation is over all nodes that are nearest neighbors ($nn$) of a given node $i$. Throughout this paper, the dynamics of the coupled system is studied for high inhibition and weak electrical coupling levels, with the value of the synaptic conductance fixed at $g_{syn}=0.81$ (unless mentioned otherwise) and varying the gap-junctional conductance $g_{gap}$. For all simulations reported in this work, we study the dynamics obtained when a sub-threshold stimulus of $I_{ext}=0.03$ is given to the pre-synaptic neuron. The equations are solved using variable step stiff solver ODE15s of \verb|MATLAB Release 2010b| with a tolerance of $1e^{-8}$ and verified the results using $4^{th}$ order Runge-kutta method .

\section*{III. Results}
%\section{Results}

%\clearpage

\begin{figure*}[htbp]
%\centering
\includegraphics[width=0.9\linewidth]{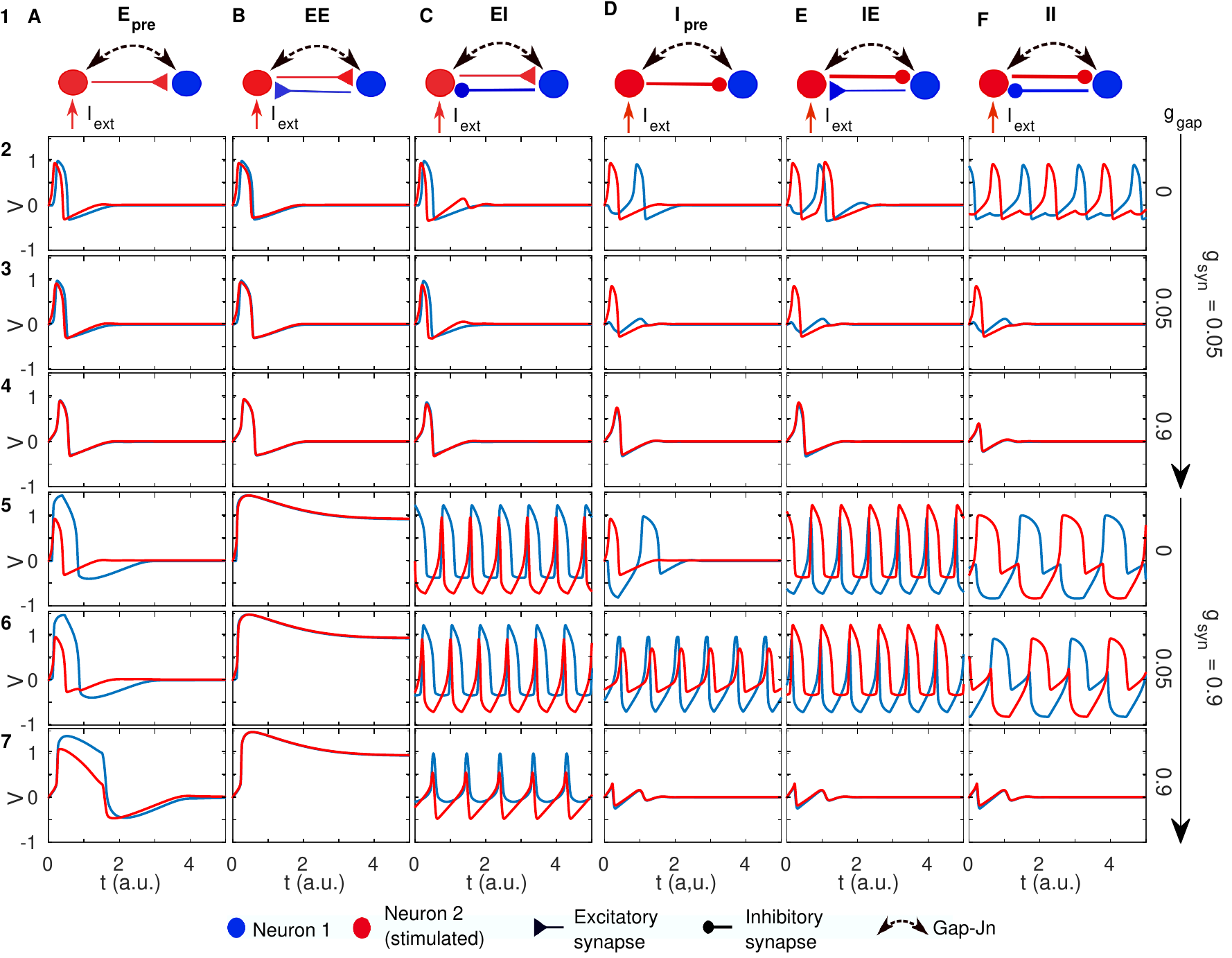}
\caption{(color online) Activity profiles of the pre- and post-synaptic neurons observed by establishing a synaptic and gap-junctional connections in distinct combinations between two neurons. 
(A1-F1) Schematic diagrams representing the possible ways of setting up a synaptic and a gap-junctional coupling between two neurons. 
The pre- and the post-synaptic neurons in the schematic are marked by red and blue color circles respectively. The six different possibilities are as follows: [L-R] (A1) uni-directional excitatory synaptic coupling along with gap-junction [$E_{pre}$], (B1) bi-directional coupling between two excitatory neurons along with gap-junction [EE], (C1) bi-directional coupling between an excitatory and an inhibitory neuron along with gap-junction [EI],
(D1) an uni-directional synapse from an inhibitory pre-synaptic neuron along with gap-junction [$I_{pre}$], (E1) bi-directional coupling between and inhibitory and an excitatory neuron along with gap-junction [IE] and (F1) bi-directional coupling between two inhibitory neurons along with gap-junction [II].  For all the aforementioned configurations, the neurons are originally in their resting states, while only the pre-synaptic neuron (excitatory or inhibitory) is subjected to a sub-threshold stimulus $I_{ext}$. The stimulated neuron along the column (A-C) is excitatory, whereas the stimulated neuron along the column (D-F) is inhibitory. 
(A2-F2, A3-F3 and A4-F4) correspond to low value of synaptic conductance $g_{syn}=0.05$, whereas panels (A5-F5, A6-F6, A7-F7) correspond to high value of synaptic conductance $g_{syn}=0.9$. Both the values of $g_{syn}$ are indicated at the right end of the figure. Along each row, the value of gap-junctional conductance $g_{gap}$ is kept constant. The arrows on the right represent the increasing direction of $g_{gap}$, whose precise values viz., $0, 0.05, 0.9$ are indicated at the end of each row corresponding to low and high value of $g_{syn}$. Note that at least one inhibitory synaptic connection is necessary for sustained oscillations in the system. The simplest possible two neuron system to show oscillations corresponds to the configuration (D1), which comprises one inhibitory synapse and a gap-junction.
}
\label{fig1}
\end{figure*}

\paragraph*{}We have carried out simulations of systems of coupled excitable neurons that exhibit Fitzhugh Nagumo dynamics, as described in the preceding section.
%\nameref{sec:model}. 
The activity profiles of both pre- and post-synaptic neurons (shown in red and blue respectively), for different high and low inhibitory synaptic $g_{syn}$ and gap-junctional $g_{gap}$ conductance are displayed in Fig.~\ref{fig1} for the simplest case of $N=2$ coupled neurons. In all of these simulations, the neurons are originally in their resting states, while the pre-synaptic neuron alone is subjected to a brief sub-threshold pulse $I_{ext}$. 

The schematic representation shown along the top row of Fig.~\ref{fig1} (A1-F1) displays six possible ways of establishing a synaptic connection (uni-directional and bi-directional) between two neurons (excitatory or inhibitory) in the presence of a gap-junction, namely: $E_{pre}$, EE, EI,  $I_{pre}$, IE and II. The neuronal activities corresponding to each of the aforementioned connectivities is shown along the columns, with their specific connections represented by a schematic diagram on top of each column. Along each row, the conductance values $g_{syn}$ and $g_{gap}$ are kept constant and their specific values are mentioned on the right side of the figure. For all our simulations on coupled neurons, we have assumed that a synaptic connection exists irrespective of the existence of a gap-junction, hence $g_{syn} > 0$ and $g_{gap} \geq 0$. The neuronal firing patterns corresponding to low and high synaptic inhibition are shown in Fig.~\ref{fig1} (A2-F2, A3-F3 and A4-F4) and Fig.~\ref{fig1} (A5-F5, A6-F6, A7-F7) respectively, for varying $g_{gap}$.

%\clearpage
\begin{figure*}[htbp]
\centering
\includegraphics[width=0.9\linewidth]{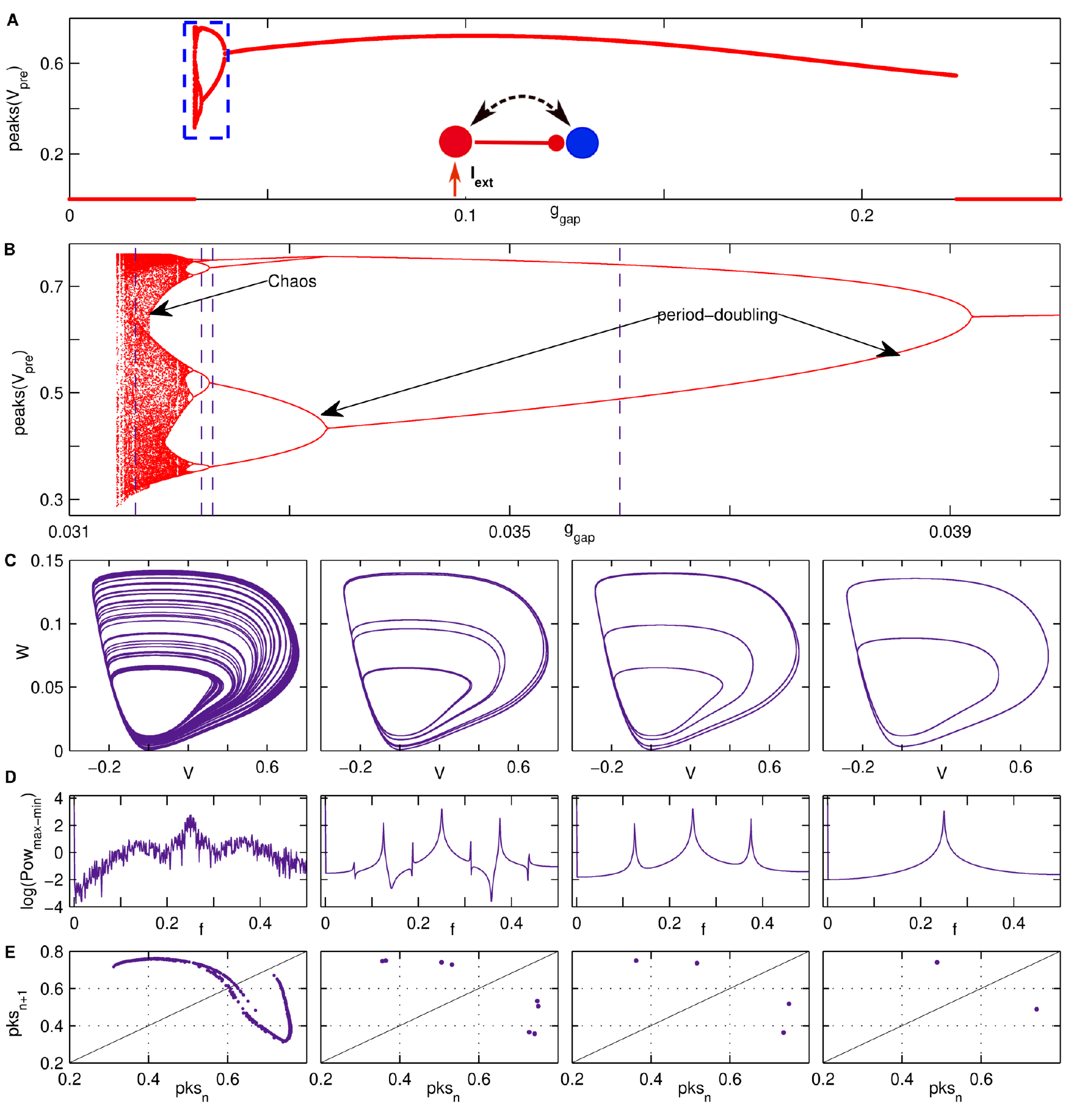}
%\begin{multicol}
\caption{(Color online) Period-doubling route to chaos in  coupled neurons connected by an inhibitory synapse and a gap-junction.
(A) The schematic (shown as an inset) represents the simplest configuration of $N=2$ coupled neurons with uni-directional inhibitory synapse and gap-junctions when the pre-synaptic neuron alone is subjected to a brief sub-threshold stimulus. The bifurcation diagram, is obtained by varying the $g_{gap}$ values (along x-axis), while $g_{syn}=0.81$ is fixed. Plotted along the y-axis are the peak values of the pre-synaptic membrane potential $V_{pre}$, obtained for the last 100 time points. The coupled neurons show oscillatory dynamics for an intermediate range of $g_{gap}$ values, and stable fixed points otherwise. (B) The enlarged portion corresponding to the blue rectangular region in panel (A), shows period-doubling route to chaos, on decreasing $g_{gap}$. (C) The phase space trajectory of the pre-synaptic neuron corresponding to those values of $g_{gap}$ indicated by violet broken lines in panel (B). [L-R] represents complex dynamics namely chaos, period-8, period-4 and period-2 oscillations respectively. (D) shows the corresponding power spectral density ($Pow_{max-min}$) of a discrete time series constructed by taking the maximum and the minimum points of the original time series of the pre-synaptic neuron and (E) shows the Poincaré map (or the return map obtained by plotting the $n^{th}$ and the $(n+1)^{th}$ peak obtained from the time-series of the pre-synaptic neuron) corresponding to the neuronal dynamics shown in panel (C).}
%\end{multicol}
\label{fig2}
\end{figure*}

It is well known that the activity patterns of the coupled neurons vary depending on (1) the type of neurons viz., excitatory or inhibitory (2) they type of synaptic coupling viz., uni-directional or bi-directional and (3) the coupling strengths $g_{syn}$ and $g_{gap}$. The precise contribution of such factors on the activity profiles are discussed in detail in Fig:~\ref{fig1}. For systems that have low synaptic inhibition and no gap-junctions (Fig.~\ref{fig1} (A2-C2)), we observe that an excitatory pre-synaptic neuron causes the firing of its post-synaptic neuron almost simultaneously. But in the case of an inhibitory pre-synaptic neuron, the post-synaptic neuron fires an action potential after being released from the suppressing effect of the inhibitory pre-synaptic neuron, which is called the post-inhibitory rebound which can be seen clearly in Fig.~\ref{fig1} (D2-E2). Owing to this post-inhibitory rebound effect of inhibitory synapses, in the presence of bi-directional inhibitory coupling, the coupled system with the II configuration exhibits sustained oscillations (Fig.~\ref{fig1} (F2)). Furthermore, introducing gap-junctional conductance $g_{gap}$ causes synchronized behavior of neurons (Fig.~\ref{fig1} (A3-F3) and (A4-F4)) which may or may not result in an action potential (depending on the pre-synaptic neuron). In the bi-directional inhibitory synapse case, where neurons originally exhibited sustained oscillations, introducing the gap-junction kills sustained activity (Fig.~\ref{fig1} (F3 and F4)).  Thus, at low synaptic conductance, the coupled system does not give rise to persistent activity in the presence of gap-junctions.   

Having studied the effects of low synaptic conductance on the neuronal firing, we now focus on how high synaptic inhibition affects the behavior of the coupled neurons. In the case of systems with no gap-junctions, we observe that when the synaptic conductance $g_{syn}$ value is high, the neurons exhibit persistent activity only for bidirectional synapses where at least one of the neurons is inhibitory (EI, IE and II as represented in Fig.~\ref{fig1} (C1, E1 and F1) respectively). These bidirectional synapses continue to sustain oscillations in the presence of weak gap-junctional conductance (Fig.~\ref{fig1} (C6, E6 and F6)) but as the gap-junctional conductance grows stronger (Fig.~\ref{fig1} (C7, E7 and F7)), IE and II can no longer sustain oscillations while EI continues to show persistent activity. The configurations EI and IE are identical, while only the stimulated neurons are distinct in both cases (excitatory for EI and inhibitory for IE). Hence both the configurations exhibit qualitatively similar firing pattern in all the cases (Fig.~\ref{fig1} (C2-C6 and E2-E6)) except for  high values of $g_{syn}$ and $g_{gap}$ (Fig.~\ref{fig1} (C7 and E7)), where EI configuration shows oscillations due to the stimulation of excitatory pre-synaptic neuron while IE configuration does not initiate oscillations as the stimulated inhibitory neuron suppresses the excitation of its post-synaptic neighbor.
Among all the configurations, there is only one uni-directional synaptic configuration which shows sustained oscillations, and that is $I_{pre}$ (Fig.~\ref{fig1} (D1)), in the presence of a strong inhibitory synapse and a weak gap-junctional conductance (Fig.~\ref{fig1} (D6)). However, at strong $g_{gap}$ (Fig.~\ref{fig1} (D7)), oscillations are not sustained. This is the simplest system of two coupled neurons, with a single synapse and a gap-junction, that exhibits persistent activity. Therefore, in order to understand the interplay between the synapse and the gap-junction in creating persistent activity in a neuronal network, we carry an in-depth exploration of the dynamical behavior of this specific configuration viz., $I_{pre}$. The above given results indicate that the interplay between synaptic and gap-junctional conductance values are crucial in determining the dynamical behavior of the system. 

%\clearpage
In a system of two neurons coupled by a strong inhibitory synapse and a gap-junction, the change in the dynamical behavior from non-oscillatory to oscillatory to non-oscillatory states at high inhibition is studied by varying the gap-junctional conductance $g_{gap}$ (see Fig.~\ref{fig2}). Starting from their initial resting states, i.e. $(V_{i},W_{i},S_{i})=(0,0,0)$ where $i = 1,2$, this coupled neuronal  system results in persistent dynamical activity, when a sub-threshold pulse is applied to the pre-synaptic inhibitory neuron. The bifurcation diagram (Fig.~\ref{fig2} (A)), with the gap-junctional conductance along the x-axis and the  membrane potential ($V_{pre}$) of the pre-synaptic neuron along the y-axis, is obtained for the case of high synaptic coupling. This bifurcation diagram shows that the neurons exhibit oscillatory behavior for a specific range of gap-junctional conductance $0.031 < g_{gap} < 0.25$, while their extreme values does not sustain oscillations. The striking feature of this minimal system of two coupled inhibitory neurons is their ability to show complex dynamical behavior, for a restricted range of $g_{gap}$ (enclosed within the blue box). This range, although very small, exhibits a rich dynamical repertoire as can be seen in the magnified view in Fig.~\ref{fig2} (B). We find that when $g_{gap}$ is reduced below a critical value ($\approx0.0392$), the system undergoes a period-doubling bifurcation thereby converting the attractor from period-1 to period-2. Further decrease in $g_{gap}$ results in the emergence of period-4 and  period-8 attractors, after which we observe the onset of chaos, which is indicated by the appearance of solid red bands formed due to the merging of successive bifurcations. We thus note that there is a minimum value of the gap-junctional conductance $g_{gap}$ below which the neurons do not show persistent activity. To verify if there exists a chaotic attractor (but with a smaller basin of attraction) below the minimum $g_{gap}$ value, we performed annealed simulations (results not shown), where we allow the neurons to reach a particular attractor and gradually vary the $g_{gap}$ using the current state as the initial condition as opposed to quenched simulations where for every $g_{gap}$ we start from resting state initial conditions, and found the existence of only the resting state attractor. This sudden disappearance of the chaotic attractor on reducing $g_{gap}$ 
can be attributed to the well studied boundary crisis~\cite{Grebogi1983}. Moreover, we show that, the dynamics exhibited by the post-synaptic neuron is qualitatively similar to that of the above discussed pre-synaptic neuron (See Supporting Information Fig:~\ref{SI2}). Thus, it is clear from the aforementioned results that a system of two neurons coupled through both chemical and electrical synapses is the minimal network that can exhibit both sustained activity as well as complex dynamics such as chaos. 

In order to further analyze the observed dynamical patterns, we consider four different $g_{gap}$ values indicated by vertical broken lines in violet. The phase space trajectory corresponding to the chosen conductance $g_{gap}$ values are shown in Fig.~\ref{fig2} (C), with complex behavior such as [L-R] chaos, period-8, period-4, period-2 oscillations. In other words, the dynamics of the coupled system changes from complex chaotic oscillations to periodic oscillations, when $g_{gap}$ is increased yet constrained to the narrow range. In order to distinguish between various complex oscillatory dynamics, Power spectral densities of the corresponding time-series are often used. In this paper, instead of the full time series, we calculate the maximum and minimum values of each oscillation (from the original time series) and construct a discrete time series. This method along with the power spectral density (PSD) of full time series was shown to be effective in distinguishing various chaotic attractors in ~\cite{Suzuki2016}. For our paper, we found that the power spectral density calculated from the maximum-minimum time series distinguishes the different types of oscillations effectively (See Supporting Information Fig:~\ref{SI1}). For period-2,4 and 8 oscillations, we observe peaks for certain frequencies alone but for chaotic dynamics, we can see spikes throughout the entire frequency range. Yet another efficient way of distinguishing various periodic and non-periodic oscillations is the Poincaré map shown in Fig.~\ref{fig2} (E), corresponding to the four different oscillatory patterns. The Poincaré map which is obtained by plotting the $n^{th}$ and the $(n+1)^{th}$ peak, which can be calculated from the neuronal time series. We find $N$ discrete points with $N = 2, 4, 8\dots$ for period-2,4,8 oscillations and this regular geometric pattern is lost when the system shows chaotic oscillations. Hence, the bifurcation diagram along with the Poincaré map and the PSD of the discrete time series indicates that the non-periodic oscillations observed are indeed chaotic.

\begin{figure}[htbp!]
\centering
\includegraphics{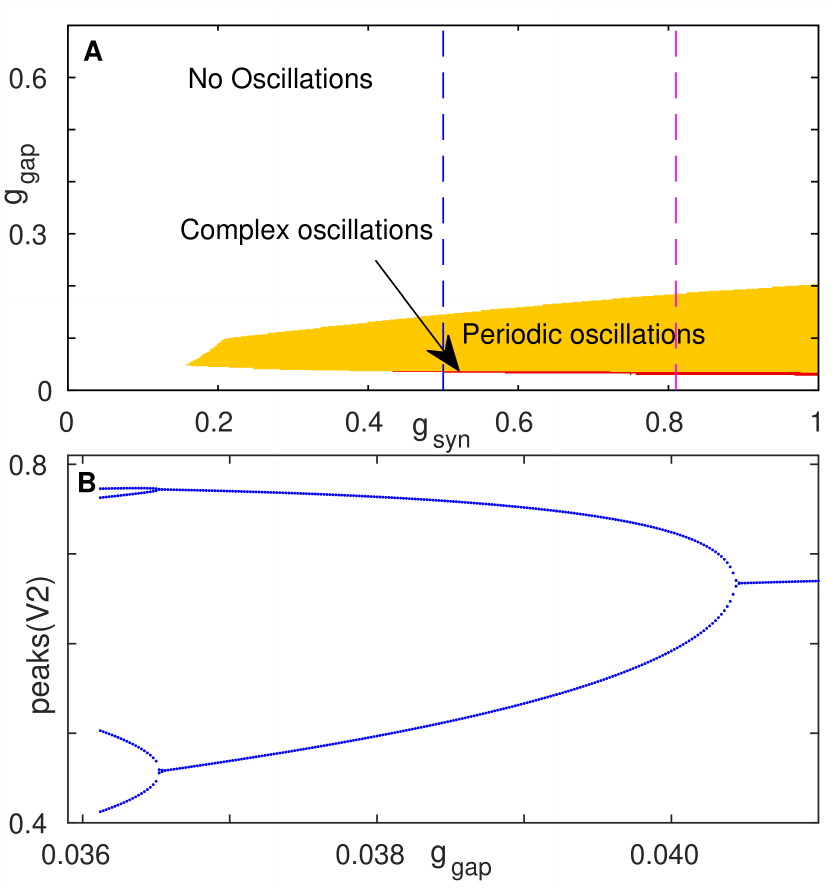}
\caption{(Color online)  Different dynamical regimes exhibited by $N=2$ coupled neurons connected by an inhibitory synapse and a gap-junction.
(A) The $(g_{syn}, g_{gap})$ parameter space, marked by the attractors to which the system converges to, viz. No Oscillations, Periodic and Complex oscillations ($>$ period-1 oscillations, indicated by an arrow), starting from their resting states, i.e. $(V_{i},W_{i},S_{i})=(0,0,0)$ where $i = 1,2$. The blue broken lines represent the $g_{syn} = 0.5$  value used for the bifurcation diagram shown in panels (B). The broken line in magenta in (A) corresponds to the $g_{syn} = 0.81$ value in Fig.~\ref{fig2} (A).
(B) The bifurcation diagram is obtained by varying the gap-junctional conductance $g_{gap}$ when a brief pulse is applied to the pre-synaptic neuron alone, fixing $g_{syn}=0.5$. Plotted along the y-axis are the peak values of the pre-synaptic membrane potential $V_{pre}$, obtained for the last 100 time points. On decreasing $g_{gap}$, the system shows premature termination of the period-doubling bifurcation (at period-4 oscillations).}
%$T=100 a.u$ 
\label{fig3}
\end{figure}

While the results mentioned above by applying a brief pulse to the pre-synaptic inhibitory neuron show many different activity patterns, we comprehensively detail the activity patterns that arise across $g_{syn},g_{gap}$ parameter space (Fig.~\ref{fig3} (A)) and identify regions of periodic (yellow region), complex oscillations (narrow black region indicated by an arrow) and no oscillations (white region) when the system is subjected to a brief pulse. We see that the complex oscillations are limited to a narrow range of weak gap-junctional $g_{gap}$ and for strong synaptic $g_{syn}$ coupling strengths. Furthermore, the range of $g_{gap}$ for which the system shows oscillations increases with $g_{syn}$ and the boundary between periodic and no oscillation regime shows a monotonic behavior and the precise shape of the boundary is attributed to the choice of initial conditions (here, the resting state values). We know from Fig.~\ref{fig2} that the coupled system gives rise to chaotic behavior for synaptic conductance as high as $g_{syn}=0.81$, when a brief pulse is applied. In order to understand if the region of complex oscillations always include chaotic behavior, we chose $g_{syn}=0.5$ for this analysis, as complex oscillations begin to appear close to this value of $g_{syn}$. The bifurcation diagram obtained by varying $g_{gap}$ (Fig.~\ref{fig3} (B)) shows the presence of complex dynamics viz., period-2, period-4 oscillations, but for a very narrow region of $g_{gap}$. Although the system undergoes a period-doubling bifurcation, their activity is prematurely terminated with lowering $g_{gap}$ and hence the system does not show chaotic behavior. This shows that the coupled system with a brief pulse can exhibit chaotic behavior but at a much higher value of $g_{syn}$.

%\clearpage
\begin{figure}[htbp!]
\centering
\includegraphics{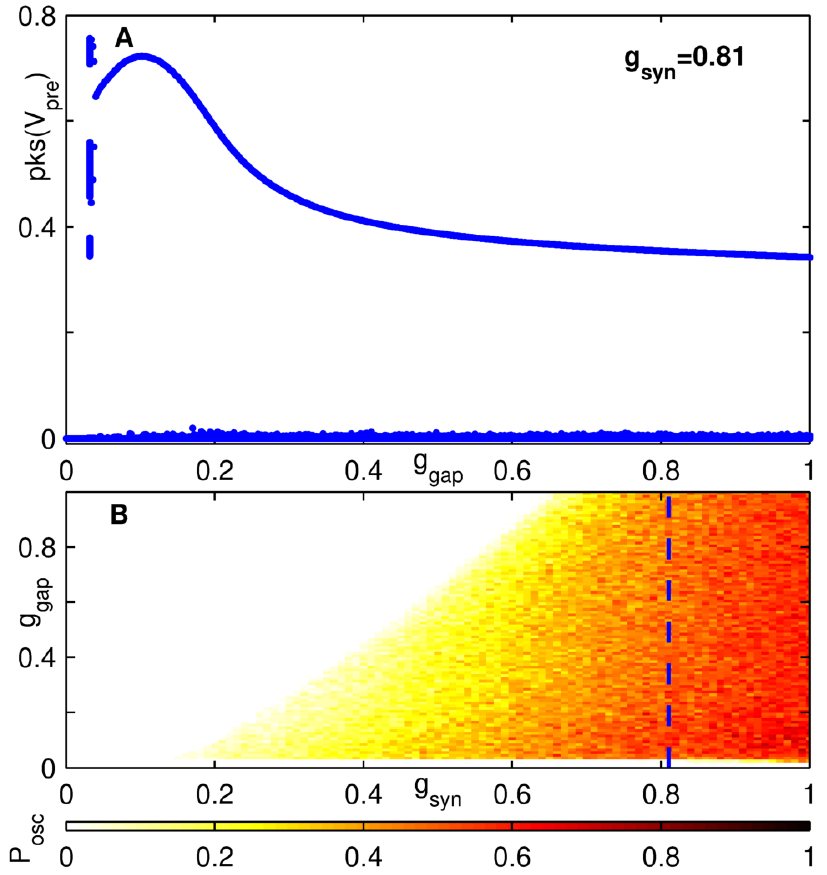}
\caption{(Color online) Bi-stability observed in $N=2$ coupled neurons connected by an inhibitory synapse and a gap-junction, starting from random initial conditions.
(A) The bifurcation diagram is plotted by varying gap-junctional conductance $g_{gap}$ values, at a fixed value of synaptic conductance $g_{syn}$. Plotted along the y-axis are the peak values of the pre-synaptic membrane potential $V_{pre}$, obtained for the last 100 time points. Starting from random initial conditions for each value of $g_{gap}$, the coupled neurons exhibit co-existence of limit cycle and fixed point attractors. Whereas an equivalent diagram (plotted at same $g_{syn}$ as (A)) shown in Fig.~\ref{fig2} (A), exhibited only one stable attractor for a given value of $g_{gap}$ owing to the resting state initial conditions of the neurons. 
(B) The probability of obtaining oscillations ($P_{osc}$) in $(g_{syn},g_{gap})$ parameter space for the coupled neurons is obtained starting from random initial conditions. $P_{osc}$ represents the basin size of limit cycle attractor and this value increases with the strength of inhibition. Note that a minimum $g_{gap}$ is required for the coupled system to show oscillations.
The blue broken line indicates the value of $g_{syn} = 0.81$ corresponding to the bifurcation diagram in (A). Both (A) and (B) are obtained for 100 random initial conditions.}
\label{fig4}
\end{figure}

\begin{figure}[htbp!]
\centering
\includegraphics{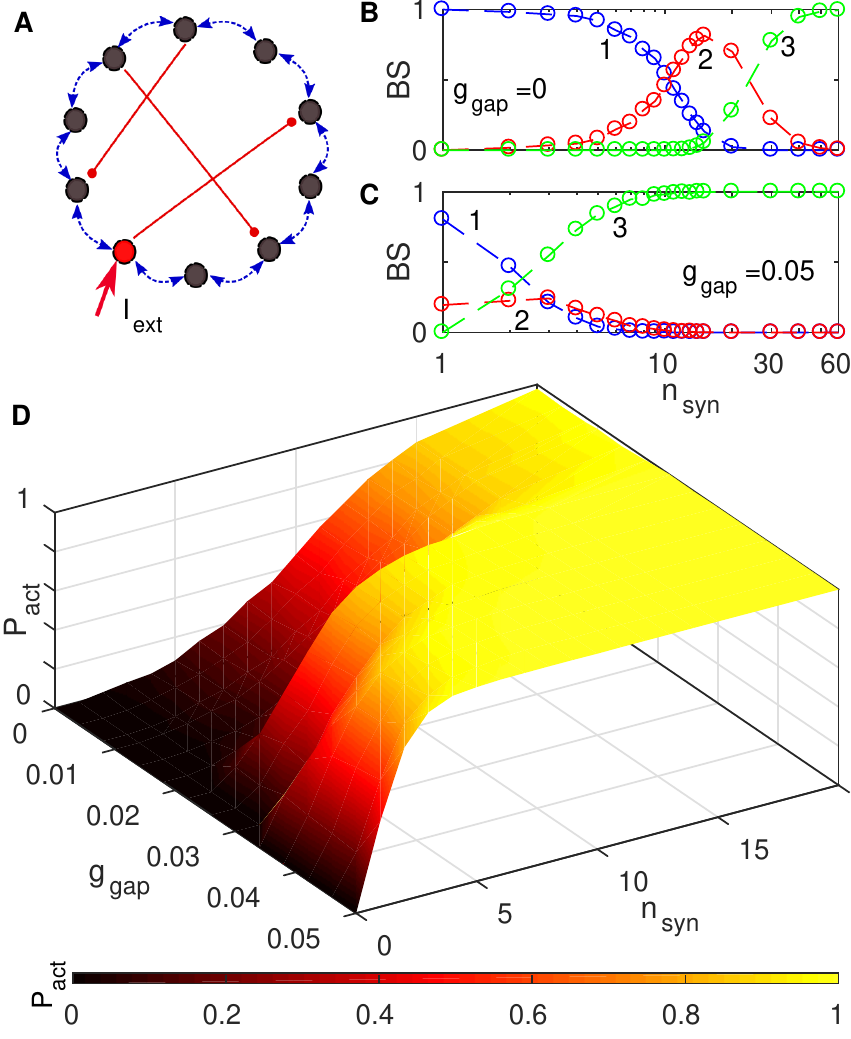}
\caption{(Color online) Persistent activity in a multiplex set-up of $N=10$ neurons coupled through synapses and gap-junctions, arranged in a one-dimensional ring.
(A) Schematic representation of a network of coupled inhibitory neurons (black circles) connected to their nearest neighbors by gap-junctions (shown in blue arrows). The red lines represent the inhibitory synaptic links between randomly chosen pairs of neurons. Only one of the many pre-synaptic neurons (marked in red) receive a short pulse ($I_{ext}$). Simulating such a system by varying the number of synaptic connections ($n_{syn}$) for different gap-junctional conductance values $g_{gap}$ can result in the convergence of the system to one or more of the following attractors namely (1) a fixed point attractor (with no oscillations) (2) a chimera state with oscillations of only few of the neurons (3) a global oscillatory state, i.e. oscillations of all the neurons in the system   
(B-C) Co-existence of multiple attractors (marked by 1, 2 and 3, whose characteristic dynamics are described above) is indicated by the variation in their basin sizes (BS) for $g_{gap}=0$ and $g_{gap}= 0.05$ respectively, where BS represents the fraction of initial conditions that result in a particular dynamical attractor.  
(D) The probability of obtaining active neurons (showing oscillations) ($P_{act}$) for different values of $n_{syn}$ (along the x-axis) and $g_{gap}$ (along the y-axis). Two different transitions occur at $g_{gap}=0.02$ and $g_{gap}=0.03$ respectively and at every transition (along increasing $g_{gap}$), a non-zero probability of activity $P_{act}$ is observed even for lower $n_{syn}$.}
\label{fig5}
\end{figure}

To quantitatively analyze the robustness of the observed oscillatory behavior and the oscillatory and non-oscillatory boundary, we considered the minimalistic network of two couped neurons with each neuron starting from a random initial state (as opposed to the special resting state initial condition used for the rest of simulations in this paper). The behavior of the system is studied by applying a brief pulse to the pre-synaptic neuron. The bifurcation diagram Fig.~\ref{fig4} (A) shows that the system either goes to the oscillatory branch (top) or stable fixed point (bottom), thereby exhibits a bi-stable behavior, starting from a random initial state. The synaptic conductance value for which the bifurcation diagram is plotted is $g_{syn}=0.81$ and the blue broken line in Fig.~\ref{fig4} (B) indicates the same. The bifurcation diagram in Fig.~\ref{fig2} (A) and Fig.~\ref{fig4} (A), although have same set of parameter values, they differ in the choice of their initial conditions. We notice that the former does not show a bi-stable behavior due to resting state initial conditions, whereas the latter with random initial conditions exhibit bi-stability. Additionally, the probability of obtaining sustained oscillations $P_{osc}$ in $(g_{syn},g_{gap})$ parameter space, is shown in Fig.~\ref{fig4} (B). The $P_{osc}$ provides information on the basin size (which gives information on fraction of initial conditions leading to a limit cycle attractor) for different conductance values. We considered 100 trials, each starting from a random initial condition and our results show that the probability of obtaining oscillations increases with the value of $g_{syn}$, provided the gap-junctional conductance is greater than a lower cut-off value, which is $g_{gap}>0.03$.

Although the activity of a pair of coupled inhibitory neurons have been analyzed in detail, it is important to extend the study to a network of inhibitory neurons and identify the parametric requirements that give rise to persistent activity in such networks. For this purpose, we consider a network of $N=10$ neurons arranged in a one-dimensional ring topology coupled through coupled through randomly chosen pairs of synapses and nearest neighbor gap-junctions and the results are summarized in Fig.~\ref{fig5}. Although the synaptic conductance $g_{syn}=0.81$ is fixed, the number of synaptic connections $n_{syn}$ in the network can be varied from a minimum of one connection to a maximum of $N(N-1)$ connections. Hence, for each value of gap-junctional conductance $g_{gap}$, the value of $n_{syn}$ is varied and the network activity is observed, by stimulating only one of the pre-synaptic neuron with a short sub-threshold stimulus $I_{ext}$. Such a system exhibits co-existence of multiple attractors viz., (1) a complete quiescent state or fixed point attractor, where the initial stimulus is not sufficient to generate oscillations (2) an intermediate state where only few of the oscillators oscillate while others are at rest, which is the characteristic of a chimera state and (3) a global oscillatory state, i.e. all the nodes (neurons) exhibit self-sustained oscillations. The Basin size (BS) (corresponding to each of the attractors mentioned above) displayed in Fig.~\ref{fig5} (B-C) is obtained for two different values of $g_{gap}=0, 0.5$ respectively, as the $n_{syn}$ is varied. In each of these panels, the fraction of initial conditions that converge to one of the attractors mentioned above are marked in blue, red and green colors respectively. What we observe is that the network with only synaptic connections ($g_{gap}=0$) as in Fig.~\ref{fig5} (B) demands as high as $n_{syn}=40$ synapses for all the neurons to show oscillations. For extremely low value of $n_{syn}$, the system does not show oscillations and later we find co-existence of fixed point attractor and a chimera state, eventually leading to global oscillatory state. On the contrary, the system with $g_{gap}=0.05$ shown in Fig.~\ref{fig5} (C), requires comparatively lesser synaptic connections $n_{syn} \approx 10$ for oscillation of all the nodes. Moreover, we see that even a single synapse can result in oscillation of at least few of the neurons taking the system directly to a chimera state, unlike the $g_{gap}=0$ case. The surface plot Fig.~\ref{fig5} (D), with varying $n_{syn}$ along the x-axis and different $g_{gap}$ along the y-axis shows the probability of obtaining activity in the network $P_{act}$ (shown along the z-axis). On increasing $g_{gap}$, we observe two distinct transitions corresponding to two different values of gap-junctional conductance, viz. $g_{gap}=0.02$ and $0.03$ respectively. At each transition (along the increasing $g_{gap}$ direction), we observe a higher probability of obtaining activity $P_{act}$ corresponding to lower $n_{syn}$. In other words, increasing $g_{gap}$ increases the $P_{act}$ even with lesser number of synaptic connections. Hence, for $g_{gap} \geq 0.03$, we obtain a non-zero probability of obtaining oscillations when compared to conductance values lower than 0.03 i.e.$g_{gap}=0.03$.  A further increase in gap-junctional conductance $g_{gap} > 0.05$ results in failure of sustained activity, mainly due to the synchronizing ability of gap-junctions. Even as we increase the system size, the results remain qualitatively same (See Supporting Information Fig:~\ref{SI5}). Hence, it is apparent that weak gap-junctional conductance $g_{gap}$ helps in achieving global oscillatory state (with all the neurons oscillating) even with minimal synaptic connections, under strong inhibition (with $g_{syn} =\ 0.81$). 

\section*{IV. Conclusion}
To conclude, we have shown how the interplay between synaptic inhibition and electrical gap-junctions results in the emergence of persistent activity.
We have analyzed various combinations of two neurons connected with synapses and gap-junctions, and we infer the following: (a) networks of excitatory neurons alone cannot exhibit persistent activity, (b) Strong inhibition is required to maintain persistent activity in the presence of gap-junctions. Our results are in agreement with~\cite{Ermentrout2006}, where it is shown that a weak synapse cannot overcome the effect of gap-junctions causing the system to converge to the stable attractor state, thereby suggesting the requirement of strong synaptic coupling to maintain persistent activity. Through our systematic investigation, we uncover the complexity involved in a minimal model of a pair of inhibitory neurons coupled through both the aforementioned synaptic modalities. We further show that this simple system undergoes a series of period-doubling bifurcations leading to chaos. Hence, our work highlights not just the combined effect of chemical and electrical synapses but also outlines the importance of inhibitory neurons in generating and maintaining persistent yet complex dynamics in networks of coupled neurons. Furthermore, our simulations on a one-dimensional ring topology gives a preliminary understanding on the role of gap-junctions in achieving persistent activity. We report here, the existence of chimera pattern that comprises neurons exhibiting both oscillatory and non-oscillatory states. Such chimera patterns have been reported in our earlier study~\cite{Janaki2019} in the context of biological pattern formation. Studies on networks of excitatory and/or inhibitory with synapse and gap-junctions have shown to give complex spatiotemporal dynamics, viz. chimera-like pattern~\cite{Mishra2017}, transient chaotic behavior~\cite{Keplinger2014} etc. Hence it would be intriguing to analyze the spatiotemporal dynamics exhibited by purely inhibitory neurons arranged in one- and two-dimensional lattices. By using a multiplex framework with nearest-neighbor gap-junctional coupling and long-range synaptic inhibition, one can study the collective dynamics exhibited by inhibitory neurons under strong synaptic inhibition. Additionally, by evolving the gap-junctional layer alone under activity-dependent plasticity~\cite{Haas2011, Pernelle2017}, keeping the synaptic layer frozen, one could potentially study the emergent behavior in a system of inhibitory neurons. Thus, our study on networks with strong inhibitory coupling when extended to large system sizes might have potential implications in maintaining working memory as they are shown to store many more patterns than their excitatory counterparts~\cite{Mongillo2018, Kim2020}. As our study consists of identical neuronal elements, extending this work to study the effects of heterogeneity could be another exciting direction of research. 
%\nonumsection{Note Added} \noindent A note can be added before
%Acknowledgments.

\begin{acknowledgments} RJ has been supported by IMSc Project of Interdisciplinary Science \& Modeling (PRISM), and the Center of Excellence in Complex Systems and Data Science, both funded by the Department of Atomic Energy, Government of India. RJ would like to thank Dr. Sitabhra Sinha and The Institute of Mathematical Sciences for the support. We also thank Dr. Rita John, Department of Theoretical Physics, University of Madras for the constant encouragement. The simulations and computations required for this work were supported by High Performance Computing facility (Nandadevi) of The Institute of Mathematical Sciences. The Nandadevi cluster is partly funded by the IT Research Academy (ITRA) under the Ministry of Electronics and Information Technology (MeitY), Government of India (ITRA-Mobile Grant No. ITRA/15(60)/DIT NMD/01). We thank Anand Pathak and Ria Ghosh for useful discussions. We also thank Shakti N.~Menon, Soumya Easwaran and K.~A.~Chandrashekar and Rishu Kumar Singh, Tanmay Mitra and Amit Sharma for their valuable suggestions in shaping the manuscript.
 \end{acknowledgments}

%====================================================================================================%
%\bibliographystyle{unsrtnat}
%\bibliographystyle{spphys}

%\bibliographystyle{plainnat}
%\bibliographystyle{jneurosci}
%\bibliographystyle{apsrmp4-1}
%\bibliographystyle{apsrev4-1}
%\bibliography{reference}


\clearpage
\newpage
%\nonumsection{Appendices} %\noindent Appendices should be used only
%when absolutely necessary. They should come immediately before
%References.
\onecolumngrid
\begin{center}
{\large {\bf SUPPLEMENTARY INFORMATION}}
\end{center}
%\appendix{}
%\clearpage
\setcounter{figure}{0}
\setcounter{equation}{0}
\renewcommand\thefigure{S\arabic{figure}}
\renewcommand\thetable{A\arabic{table}}
\renewcommand{\thesection}{\Roman{section}} 
\renewcommand{\thesubsection}{\thesection.\Alph{subsection}}

\begin{figure}[htbp!]
\begin{center}
\includegraphics{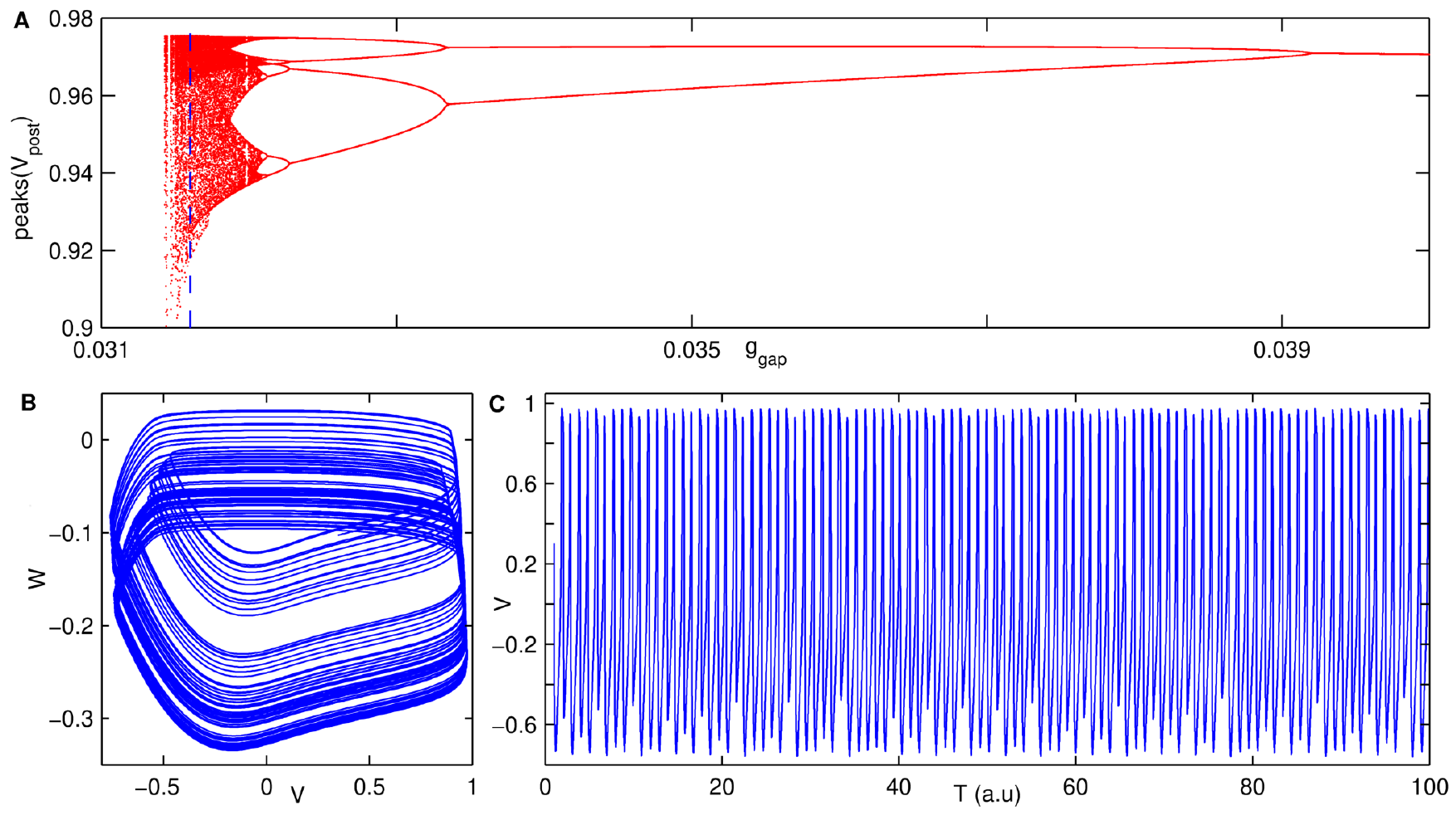}
\end{center}
%\vspace{-1.5em}
\caption{ Period-doubling route to chaos in coupled neurons with uni-directional inhibitory synapse and gap-junctions.
(A) The dynamics of the post-synaptic neuron when the pre-synaptic neuron alone is subjected to a brief sub-threshold input pulse $I_{ext}$ is shown using a bifurcation diagram. The x-axis shows varying gap-junctional conductance $g_{gap}$ values, while the synaptic conductance is fixed at $g_{syn}=0.81$. Plotted along the y-axis are the peak values of the post-synaptic membrane potential $V_{post}$, obtained for the last 100 time points. (B)-(C) represent the phase space trajectory and the time series respectively of the post-synaptic neuron corresponding to the value of $g_{gap}$ indicated by blue broken line in panel (A). This figure can be compared to Fig:~\ref{fig2} of main text, which shows the dynamics of the pre-synaptic neuron.  
}
\label{SI2}
\end{figure}

\begin{figure}[htbp!]
\begin{center}
\includegraphics{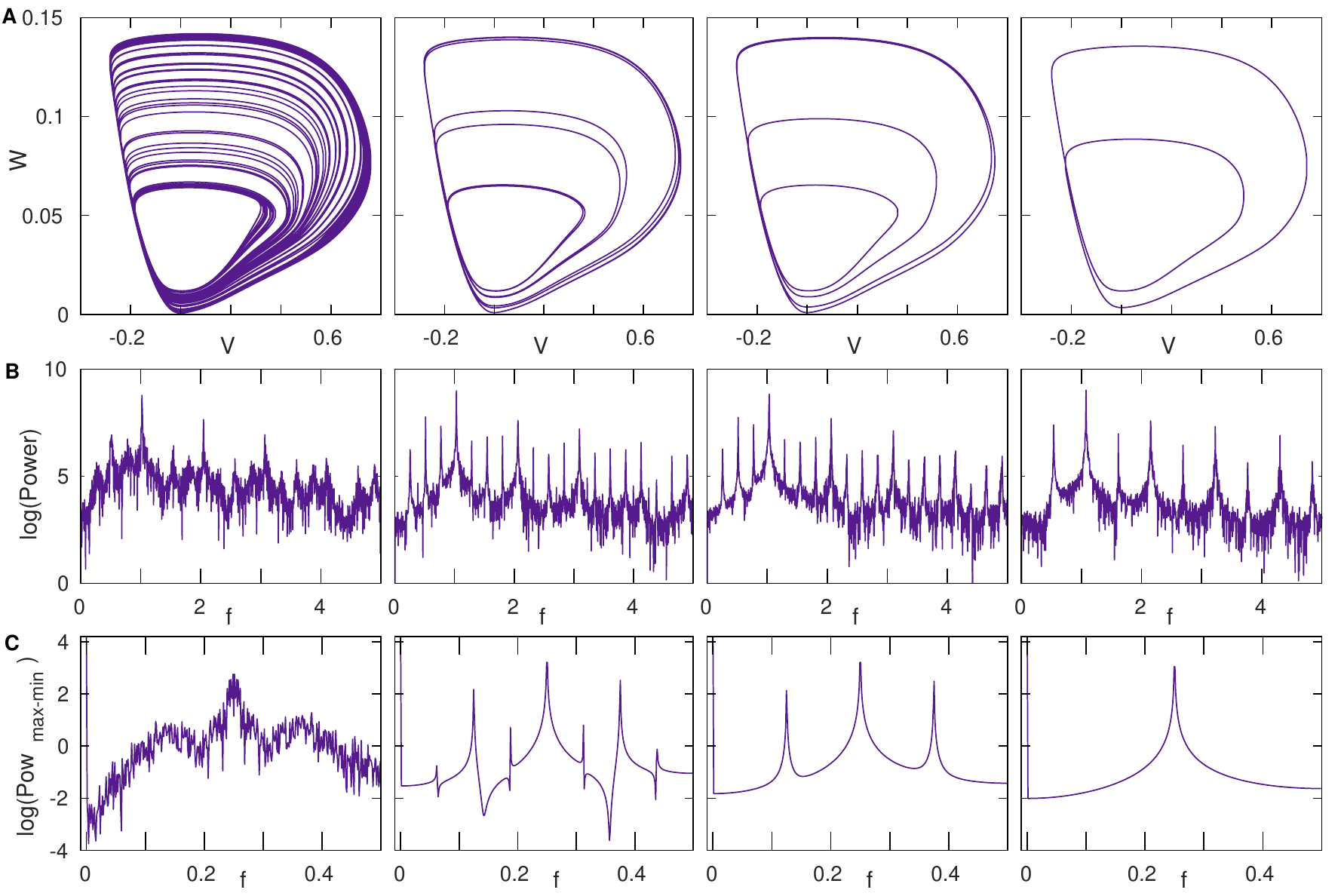}
\end{center}
%\vspace{-1.5em}
\caption{Complex dynamical behavior arising from the interaction of two coupled neurons through uni-directional synapse and gap-junction.
(A) The phase space trajectory of the pre-synaptic neuron corresponding to those values of $g_{gap}$ indicated by violet broken lines in Fig:~\ref{fig2} (B) of the main text. [L-R] represents complex dynamics namely chaos, period-8, period-4 and period-2 oscillations respectively as the $g_{gap}$ is increased. (B)-(C) shows the corresponding power spectral density (Pow) of the full time series and the ($Pow_{max-min}$) of a discrete time series constructed by taking the maximum and the minimum points of the original time series of the pre-synaptic neuron respectively. 
}
\label{SI1}
\end{figure}

\begin{figure}[htbp!]
%\centering
\includegraphics{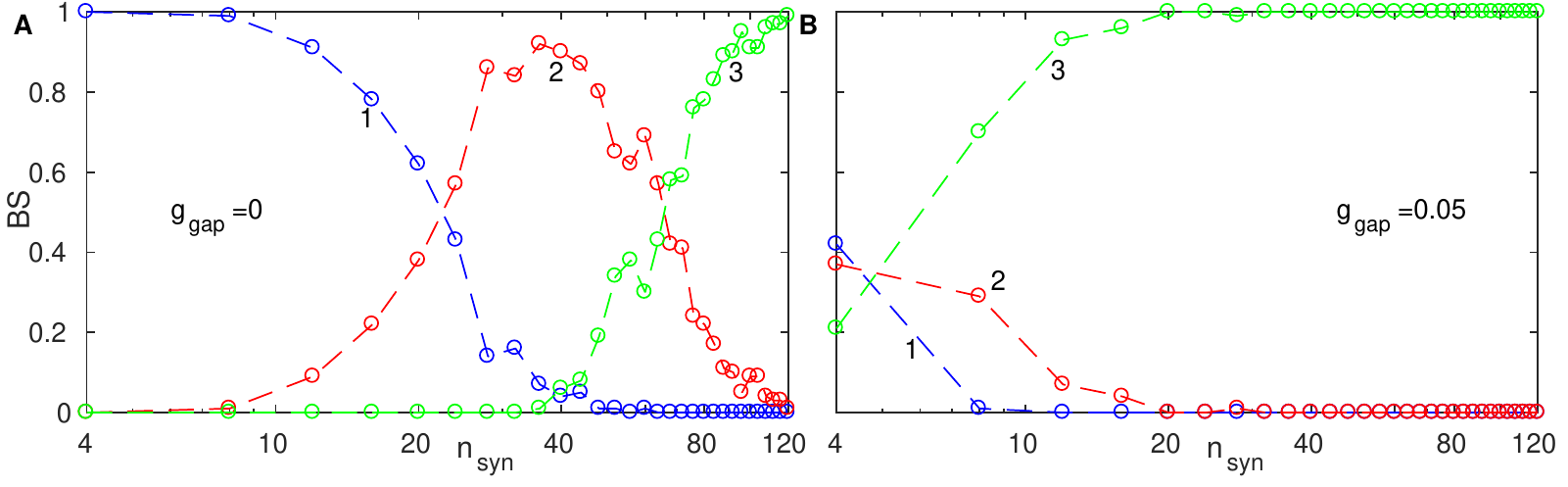}
\caption{(Color online) Persistent activity in a multiplex set-up of $N=20$ neurons coupled through synapses and gap-junctions, arranged in a one-dimensional ring.
(A-B) Co-existence of multiple attractors (marked by 1, 2 and 3, whose characteristic dynamics are described as follows: (1) a fixed point attractor (with no oscillations) (2) a chimera state with oscillations of only few of the neurons (3) a global oscillatory state, i.e. oscillations of all the neurons in the system) is indicated by the variation in their basin sizes (BS) for $g_{gap}=0$ and $g_{gap}= 0.05$ respectively, where BS represents the fraction of initial conditions that result in a particular dynamical attractor. We notice that these results for $N=20$ neurons are qualitatively similar to that of $N=10$ nodes shown in Fig:~\ref{fig5} (B-C) of the main text.
}
\label{SI5}
\end{figure}

\end{document}